\documentclass[conference]{IEEEtran}
\IEEEoverridecommandlockouts
\usepackage[ruled,vlined]{algorithm2e}
\usepackage{subcaption}
\usepackage{bbm}
\usepackage{adjustbox}
\usepackage{lipsum}
\usepackage{stmaryrd}
\usepackage{bm}
\usepackage{cite}
\usepackage{amsmath,amssymb,amsfonts}
\usepackage{algorithmic}
\usepackage{graphicx}
\usepackage{siunitx}
\usepackage{textcomp}
\usepackage{xcolor}
\def\BibTeX{{\rm B\kern-.05em{\sc i\kern-.025em b}\kern-.08em
    T\kern-.1667em\lower.7ex\hbox{E}\kern-.125emX}}
\begin{document}

    \title{\huge Connectivity-Aware Task Offloading for Remote Northern Regions: a Hybrid LEO-MEO Architecture\\
    \thanks{ This work was supported in part by Fonds de recherche du Québec secteur Nature et technologies (FQRNT) and the Natural Sciences and in part by Engineering Research Council (NSERC) of Canada Alliance grant ALLRP~579869-22 ("Artificial Intelligence Enabled Harmonious Wireless Coexistence for 3D Networks (3D-HARMONY)"). }}

\author{Mohammed Almekhlafi$^*$, Antoine Lesage-Landry$^*$$^\ddag$, Gunes Karabulut Kurt$^*$\\
$^*$Department of Electrical Engineering, Polytechnique Montréal \& Poly-Grames Research Centre, QC, Canada
\\
$^\ddag$Department of Electrical Engineering, Polytechnique Montréal, Mila \& GERAD, Montréal, QC, Canada \\
Emails:~$\{$$\text{mohammed.al-mekhlafi, antoine.lesage-landry, gunes.kurt}\}$@polymtl.ca  
}

\maketitle
\IEEEpeerreviewmaketitle

\begin{abstract}
Arctic regions, such as northern Canada, face significant challenges in achieving consistent connectivity and low-latency computing services due to the sparse coverage of Low Earth Orbit (LEO) satellites. To enhance service reliability in remote areas, this paper proposes a hybrid satellite architecture for task offloading that combines Medium Earth Orbit (MEO) and LEO satellites. We develop an optimization framework to maximize task offloading admission rate while balancing the energy consumption and delay requirements. Accounting for satellite visibility and limited computing resources, our approach integrates dynamic path selection with frequency and computational resource allocation. Because the formulated problem is NP-hard, we reformulate it into a mixed-integer convex form using disjunctive constraints and convex relaxation techniques, enabling efficient use of off-the-shelf optimization solvers. Simulation results show that, compared to a standalone LEO network, the proposed hybrid LEO-MEO architecture improves the task admission rate by 15\% and reduces the average delay by 12\%. These findings highlight the architecture’s potential to enhance connectivity and user experience in remote Arctic areas.
\end{abstract}
\section{Introduction}
With the growing demand for global and seamless connectivity, non-terrestrial networks (NTNs) are anticipated to play a key role in extending coverage to remote, rural, and underserved regions \cite{10250790,10409745, 9275613, 10355086}. This is reinforced by the broad coverage provided by various satellite types within NTNs, including geostationary orbit (GEO), medium Earth orbit (MEO), and low Earth orbit (LEO) satellites \cite{10250790}. The next-generation satellites, particularly those in LEOs and MEOs, offer significantly higher throughput and lower latency compared to traditional systems, thereby enabling support for a wide range of 6G applications. As a result, NTNs are becoming a crucial component in realizing the vision of future 6G networks \cite{9275613}. In line with this vision, the 3rd Generation Partnership Project (3GPP) has been actively engaged in standardizing NTNs, with a focus on aspects such as the integration of onboard mobile edge computing (MEC) capabilities \cite{10355086}. 

\begin{figure}[t!]
\hspace{.3 in}
	\centering	\center{\includegraphics[width=.85\columnwidth,draft=false]{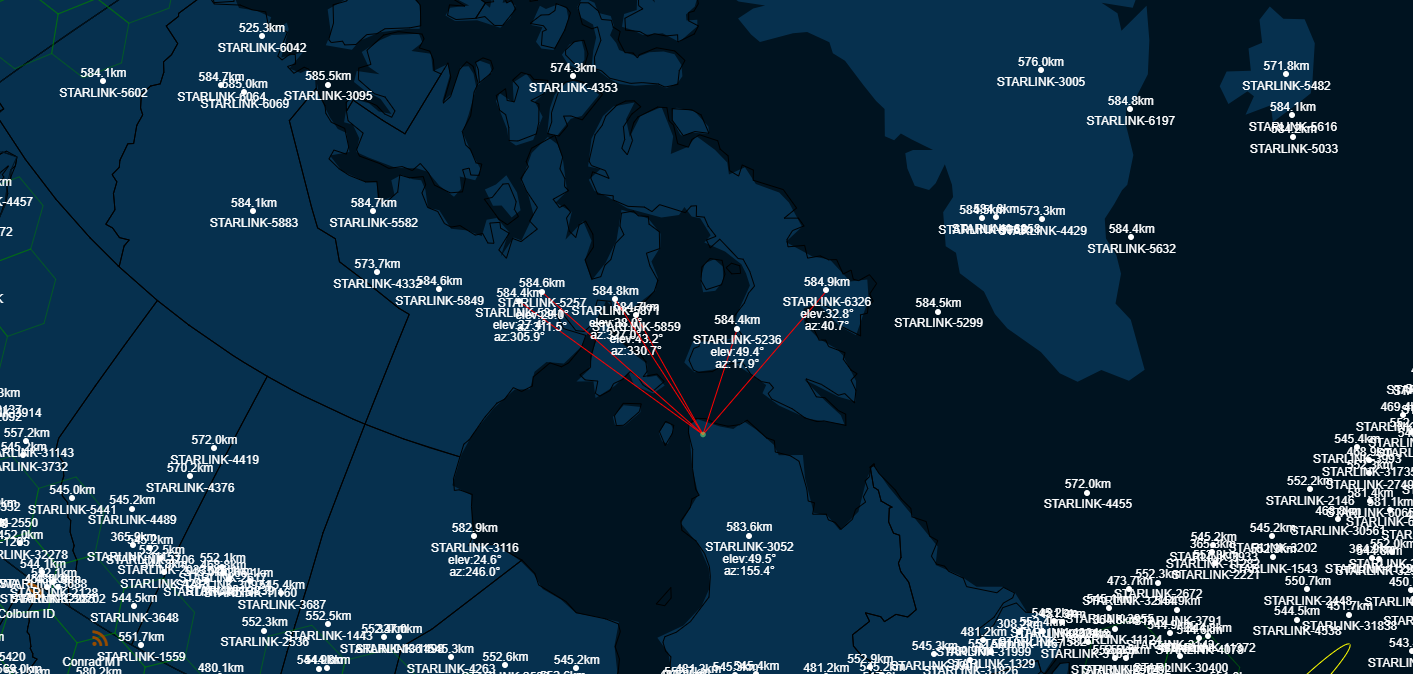}}
	\caption{Starlink network: Northern Canada. Adapted from  \cite{satellitemap}.}
	\label{fig}
    \vspace{-.3 in}
\end{figure}

MEC has demonstrated its effectiveness in terrestrial communication by significantly reducing end-to-end latency by allocating computational resources closer to the network edge \cite{8016573,9321493}. This proximity enables real-time processing for users, particularly those with limited computational capabilities, and is especially beneficial for time-sensitive or resource-intensive tasks. Building on advancements in terrestrial mobile edge computing, several studies have explored the potential of LEO satellites for edge computing. Although LEO satellites present considerable benefits for task offloading, their deployment is constrained in certain areas, such as northern Canada, where networks like Starlink are sparsely distributed and insufficient to provide comprehensive coverage, as illustrated in Fig.~\ref{fig}. These challenges are further intensified by cost considerations, as the low population density in these regions makes large-scale deployments economically unfeasible. In such cases, limited coverage and frequent outages can undermine efforts to provide reliable communication, which has been identified as a basic human right \cite{UN_HRC_Internet_2016}. Also, many studies have highlighted access inequality in remote areas, such as northern Canada, calling for effective solutions to address this issue \cite{11021288}. 

Given the limitations of LEO satellites in remote Arctic areas, this work explores the relationship between connectivity and computational efficiency in hybrid LEO-MEO networks. With broader coverage and moderate latency, MEO satellites can help mitigate LEO network limitations. In other words, MEO satellites have a longer visibility period than LEO satellites, which helps maintain continuity of service despite their higher propagation delay. The primary contributions of this work are as follows
\begin{itemize}
    \item We studied task offloading in hybrid LEO-MEO satellite networks. To achieve this, we formulate an optimization problem that maximizes the number of offloaded tasks while balancing the total delay and energy consumption, ensuring QoS requirements are met. We consider optimizing computational and frequency resources while incorporating a dynamic round-trip migration strategy. 
    \item Because the problem is NP-hard, we employ disjunctive constraints and convex relaxations to transform it into a mixed-integer convex problem that can be solved using standard optimization tools.
    \item We demonstrate the benefits of the proposed hybrid LEO-MEO scheme through extensive numerical simulations and using a realistic Starlink network model combined with an MEO satellite orbital configuration. Our results show that the proposed hybrid LEO-MEO scheme enhances user experience by increasing the task admission rate by up to $15\%$ and reducing the average delay by up to $12\%$ compared to the baseline.
    \end{itemize}   
    
\section{Related Works}
Task offloading has been extensively studied in both terrestrial and aerial networks across various applications and scenarios \cite{10613830,9772280,10077418,9693227,9725258,10620850,8314696,10103832}. The authors of \cite{10077418} introduce UAVs to address the overloading issue in vehicular edge computing networks. The formulated problem aims to minimize the average task delay while being constrained by the energy consumption of unmanned aerial vehicles (UAVs). In \cite{9693227}, self-learning solutions are proposed to decide whether the task should be offloaded to the edge or executed by the UAVs to balance between the power consumption of the vehicles and task delays. Similar problem and distributed learning are considered in \cite{9725258} for the multi-cell and multi-UAV setup. 

Ren \textit{et al.} \cite{9772280} focuses on reducing system delay by integrating HAPS while optimizing key factors such as task offloading, service caching, and computational resource distribution. They utilize the value decomposition network (VDN) method to develop an effective decision-making process. Additionally, the authors of \cite{10613830} propose two multi-connectivity offloading solutions to improve the probability of timely task completion by efficiently distributing tasks between high-altitude platform stations (HAPS) and UAVs. A task partitioning scheme is introduced in \cite{10103832}, where the task is divided into three portions and processed in parallel at the vehicle, roadside units, and a HAPS. Because the problem is not convex, a variable
replacement and the successive convex approximation method are used to solve the problem.

Task offloading optimization for NTNs is studied in \cite{10185626, 10685456,9367268,10566891,10461425,10439163}. The authors of \cite{10185626} introduce a general framework for LEO satellite-assisted task offloading, targeting real-time, high-resolution Earth observations. They suggest an iterative optimization approach, focusing on the optimal selection of data allocation, compression ratio, and processing frequency. The work in \cite{9367268} explores distributed task offloading optimization within multi-agent networks, aiming to maximize the successful offloading rate of agents. The problem is reformulated into an unconstrained optimization problem, and game theory is applied to solve it. In \cite{10566891}, the aggregation latency in a multi-hop scheme is studied where the tasks can be performed at the cloud or the LEO network. A multi-layered LEO constellation combined with terrestrial components is considered in \cite{10461425}, where the authors formulate a network selection problem to minimize latency costs. In \cite{10439163}, duelling double deep-Q-learning (D3QN) algorithm is proposed to reduce service delay for task migration in LEO networks. In \cite{10685456}, a multi-agent reinforcement learning scheme is proposed for task offloading in the Internet of Remote Things by integrating double deep Q-network (DDQN) and deep deterministic policy gradient (DDPG) to enhance decision-making efficiency and offloading performance. The connectivity challenges in Arctic regions, particularly through the use of MEO satellites, remain unaddressed. Hence, there is a clear gap in studying the benefits of integrating MEO satellites with existing LEO networks for task offloading, which is the focus of this paper.
\section{System Model}
\begin{figure}[t!]
\hspace{.3 in}
	\centering	\center{\includegraphics[width=.9\columnwidth,draft=false]{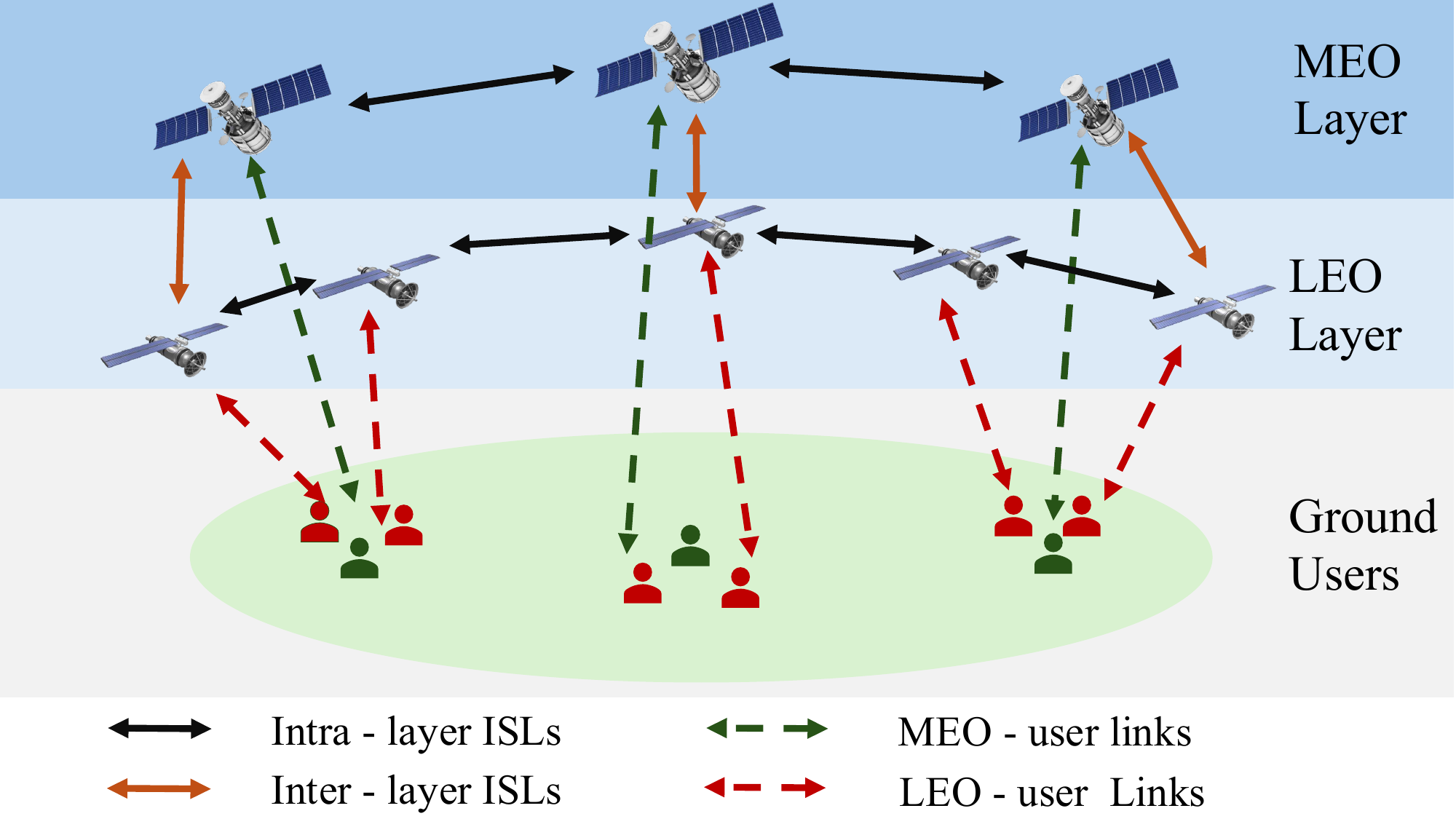}}
	\caption{Multi-layer task offloading system model. }
	\label{fig:SYSTEM_MODEL}
 \vspace{-.3 in}
\end{figure}
We consider sets of LEO and MEO satellites, denoted by \(\mathcal{L}= \{1, 2, \dots, {\rm L}\}\) and \(\mathcal{N}= \{1, 2, \dots, {\rm N}\}\), respectively, where \({\rm L} \in \mathbb{N}\) and \({\rm N}\in \mathbb{N}\) is the total number of LEO and MEO satellites, respectively. We assume that these satellites are moving within fixed orbits. The satellites provide communication services to a set of users, denoted by \(\mathcal{U}= \{1, 2, \dots, {\rm U}\}\), where ${\rm U} \in \mathbb{N}$ is the number of users located in an isolated region without access to terrestrial networks, such as Northern Canada which we consider in this work. We assume that these satellites are equipped with MEC servers, making them capable of handling computational tasks generated by ground users.
Moreover, the satellites can communicate through inter-satellite links (ISL), and offloaded tasks can be relayed to their neighbouring LEO or MEO counterparts. In this context, we model the satellite network as the graph $\mathcal{G}\left( {\mathcal{S}, \mathcal{E}}\right)$, where $\mathcal{S} \triangleq \mathcal{N}\cup \mathcal{L}  = \{1,2,\dots, \rm{L}+ \rm{N}\}$ is the set of nodes representing the satellites and $\mathcal{E}$ is the set of edges representing the ISL links between satellites. Then, if the computation resources of a satellite are insufficient to process a task, it can be forwarded to a neighbouring LEO or MEO satellite. Likewise, if a user moves out of the assigned satellite's coverage, the computation results are sent to a nearby LEO or MEO satellite for continued coverage.

For mathematical convenience, let \(\mathcal{K}= \{1,2,\dots, \rm{K}\}\) be the set of shortest paths, i.e., multi-hop routing, between all network node pairs. Each path is modelled as a tuple \(\{d_{k}, s_k, \hat{s}_k\}\), where \(d_{k}\) is the path distance in meters, \(s_k \in \mathcal{N}\) is the source satellite, and \(\hat{s}_k \in \mathcal{N}\) is the destination satellite. Without loss of generality, the frequency domain is divided into resource blocks each of bandwidth $W$. Let \({\rm B}_s\) denote the number of frequency resources of satellite  \(s \in \mathcal{S}\). For user \(u \in \mathcal{U}\), the task is modelled as a tuple \(\Omega_u = \{\zeta_u, X_u, \tau_u^{\rm max}\}\), where \(\zeta_u\), \(X_u\), and \(\tau_u^{\rm max}\) represent the packet size in bits, the required computation resources to complete the task in cycles/second, and the time needed to complete the task, respectively. Because satellites are moving at high speeds, each satellite is visible for a short period based on its speed and elevation angle. To account for orbital dynamics, we define the visibility duration of satellite \(s\) for user \(u\) as a tuple \(v_{u,s} = \{v_{u,s}^{\text{start}}, v_{u,s}^{\text{end}}\}\), where \(v_{u,s}^{\text{start}}\) and \(v_{u,s}^{\text{end}}\) are the start and end times of the visibility period.
\subsection{Routing Model}
Due to satellites' mobility, users may occasionally move out of the visibility range of the satellite they are initially associated with. Consequently, both the task itself and its corresponding computation results may need to be routed through alternative paths to ensure continuity of service. This dynamic network topology results in satellites changing positions and visibility and necessitates a flexible routing mechanism.
To address this, we introduce two optimization variables that enable effective routing management for task offloading and the subsequent delivery of computation results. The first variable controls the task offloading path (OP), ensuring that tasks are directed to a satellite capable of handling the computational load. The second variable is dedicated to computation results forwarding path (FP), ensuring the results reach the user even if the initial satellite association is no longer visible. 

Let \(i_{u,k}\) represent the offloading decision for each user \(u\) along a specific path \(k\). Specifically, \(i_{u,k} = 1\) if the task generated by  user \(u\) is offloaded through path \(k\); otherwise, \(i_{u,k} = 0\). When \(i_{u,k} = 1\), user \(u\) is thereby associated with satellite \(s_k\) as the communication node, while the computational processing of the task is conducted at the destination satellite \(\hat{s}_k\). Similarly, we introduce the binary variable \(y_{u,k}\) to control the routing of the computation results. Here, \(y_{u,k} = 1\) indicates that the results for user \(u\) are forwarded along path $k$, while \(y_{u,k} = 0\) implies that path \(k\) is not used for result forwarding. Because each ground user uses only a single OP and a single FP, the following constraints must be satisfied: 
\begin{equation}\label{const:offloading_i}
 \sum\nolimits_{k \in {\mathcal{K}}} i_{u,k} \leq 1 \, , \forall u \in {\mathcal{U}},
  \vspace{-.05in}
\end{equation}
\begin{equation}\label{const:offloading_y}
 \sum\nolimits_{k \in {\mathcal{K}}} y_{u,k} \leq 1 \, , \forall u \in {\mathcal{U}}.
\end{equation}
To ensure a closed-loop path, whereby the task is forwarded from the destination satellite of the OP to the user through FP, the following constraint must hold:
\begin{equation}\label{const:offloading_closed_path}
 \sum\nolimits_{\hat{k} \in {\mathcal{K}}} \mathbbm{I}_{s_{\hat{k}} = \hat{s}_k}\, y_{u,k} \geq i_{u,k} \, , \forall u \in {\mathcal{U}}, \forall k \in {\mathcal{K}},
 \vspace{-.05in}
\end{equation}
where \(\mathbbm{I}_{x_1 = x_2}\) is the indicator function, such that $\mathbbm{I}_{x_1 = x_2} = 1$ if the subscript statement holds true, and $0$ otherwise.

\subsection{Communication Model}
We consider that the satellites and the users are equipped with a single antenna, and the communication links between users and satellites are orthogonal. As a result, frequency resources are split orthogonally and allocated to each user. Accordingly, let \(b_{u,k,s_k}\) represent the fraction of frequency resources allocated to the \(u\)-th user by satellite \(s_k\), where \(0 \leq b_{u,k,s_k} \leq 1\). To ensure the allocated frequency resources do not exceed the capacity of each satellite, the following constraint must hold:
\begin{equation} \label{const:Frequency_f_sum}
    \sum\nolimits_{u\in \mathcal{U}} \sum\nolimits_{k\in \mathcal{K}} b_{u,k,s_k}\le 1,\, \forall s_k \in \mathcal{S}.
    \vspace{-.03in}
\end{equation}
Because the frequency allocation variable is linked with the OP selection variable, the following constraint must be satisfied:
\begin{equation} \label{const:i_b_relation}
      i_{u,k} \geq b_{u,k, s_k} \, ,\forall u \in \mathcal{U},\forall k \in \mathcal{K}, \forall s_k \in \mathcal{S}.
\end{equation}
 The transmission rate of the \(u\)-th user through path \(k \in \mathcal{K}\) is expressed as \cite{9939157}:
\begin{equation}\label{cal:achievable_rate}
R_{u,k} = b_{u,k,s_k} \,{\rm{B}}_{s_k}\,{\rm r}_{u,s_k},\, \forall u \in \mathcal{U},
\end{equation}
where ${\rm r}_{u,s_k} =  W\,  \log_2\left(1+\frac{p_u\,h_{u,s_k} }{{ \sigma}_0^2}\right)
$.
Here, \(h_{u,s_k}\) denotes the channel gain between the \(u\)-th user and satellite \(s_k\), \(p_u\) denotes the transmit power of the user, and \(\sigma_0^2\) represents the Gaussian noise power \cite{9515574}.
\vspace{-.03 in}
\subsection{Computational Model}
The task is computed at the destination satellite of the forward path. Then, let $f_{u,k,\hat{s}_k}$ be the portion of computing resources of satellite $\hat{s}_k$ allocated to task $\Omega_u$. It should satisfy 
\begin{equation}\label{const:Computing_sum}
\Bar{{\rm F}}_s+\sum\nolimits_{u\in \mathcal{U}} \sum\nolimits_{k\in \mathcal{U}} f_{u,k, \hat{s}_k}\,\mathbbm{I}_{s_{\hat{k}} =s}\le 1,\, \forall s\in \mathcal{S},    
\end{equation}
\begin{equation}\label{const:computing_relation}
I_{u,k} \ge f_{u,k, \hat{s}_k} ,\, \forall u\in \mathcal{U}, \forall k\in\mathcal{K},   
\end{equation}
where $\Bar{{\rm F}}_s$ is the portion of occupied resources. Constraint \eqref{const:computing_relation} guarantees that no computing resources are allocated if the OP~$k$ is not used to forward the computing task.
\subsection{Delay and Energy Model}
We now delve into the various sources of delay encountered in the proposed computing and communication models, which are summarized as follows:
\paragraph{\textbf{Transmission delay}} ISLs provide high transmission rates. The resulting ISL transmission delays are negligible compared to other delay components; hence, it can be ignored in the analysis \cite{10561557, 10685456}. Furthermore, the size of the computation results of the tasks is assumed to be small, allowing us to neglect their transmission delays as well. Therefore, we focus exclusively on the time required to transmit the task from the ground user to the satellite. For path $k$, the transmission time $\tau^{\rm{tr}}_{u,k}$ and the transmit energy consumption $ \rho^{\rm{tr}}_{u,k }$ of user $u$ are:
\begin{equation}\label{eq:transmission_time} 
\vspace{-.0 in}
\hspace{-.6in}
\tau^{\rm{tr}}_{u,k}= \frac{\zeta_u}{R_{u,k}}, \forall u\in \mathcal{U}, \forall k\in \mathcal{K},\vspace{-.03in}
    \end{equation}  
    \begin{equation} 
\hspace{-.3in}\rho^{\rm{tr}}_{u,k }=\frac{p_u\,\zeta_u \,i_{u,k}}{\,{\rm r}_{u,s_k}} , \forall u\in \mathcal{U}, \forall k\in \mathcal{K}.\vspace{-.05in}
    \end{equation}   
\paragraph{\textbf{Task computation delay}}  
 Offloaded tasks can be processed at the OP destination satellite. For the task \(\Omega_u\) of user \(u \in \mathcal{U}\), the computational time, $\tau^{\rm{comp}}_{u,k}$, and energy consumption, $\rho^{\rm{comp}}_{u,k}$, can be expressed as:
\begin{equation}  \label{eq:computation_time}    \hspace{-.3in}\tau^{\rm{comp}}_{u,k}=  \frac{X_u}{f_{u,k,\hat{s}_k}\, \rm{F}_s}, \forall u\in \mathcal{U}, \forall k\in \mathcal{K}, 
\vspace{-.01in}
\end{equation}

\begin{equation} 
\rho^{\rm{comp}}_{u,k}= \kappa\, X_u f_{u,k,\hat{s}_k}^2 \rm{F}_{s_k}^2, \forall u\in \mathcal{U}, \forall k\in \mathcal{K},
    \end{equation}
where $\kappa$  is the effective switch capacitance that depends on
the chip architecture \cite{10185626}.
\paragraph{\textbf{Propagation delay}} Such delay occurs when a task or its results need to be forwarded to a neighbouring satellite due to one of the following reasons: (1) insufficient computing resources on the associated satellite and/or (2) user moving out of the coverage area because of the mobility of LEO or MEO satellites. Thus, the propagation delay is composed of the task propagation delay from the user to the associated satellite plus OP delay, $ \tau^{\rm{prop,OP}}_{u,k}$, the delay of forwarding the task computational results to the end-user, $\tau^{\rm{prop,FP}}_{u,k} $, defined as:
\begin{equation}
    \tau^{\rm{prop,OP}}_{u,k} =  i_{u,k}\frac{ d_{u, s_k}+d_{k} }{c},  \forall u\in \mathcal{U},  \forall k\in \mathcal{K},
\end{equation}
\begin{equation}
    \tau^{\rm{prop,FP}}_{u,k} = y_{u,k} \frac{  d_{ \hat{s}_k,u}+d_{k}}{c},  \forall u\in \mathcal{U},  \forall k\in \mathcal{K},
\end{equation}
where $c$ is the speed of light. 

Taking into account all delay components, the total delay of a task $\Omega_u$ should satisfy the delay constraint as follows  
\begin{equation}\label{const:Delay}
     \resizebox{.88\hsize}{!}{$\sum_{k \in \mathcal{K}} \tau^{\rm{tr}}_{u,k}
    + \tau^{\rm{prop,OP}}_{u,k}
    + \tau^{\rm{prop,FP}}_{u,k}
    + \tau^{\rm{comp}}_{u,k} 
     \leq \tau_u^{\rm max} , \,\forall u \in {\mathcal{U}}.$}
\end{equation}
 Then, to guarantee task completion, the users should be visible to satellites during uplink and downlink transmissions. This is ensured through:
\begin{equation}\label{const:visibility_1}
\resizebox{.85\hsize}{!}{$v_{u,s_k}^{{\rm end}}\,i_{u,k} \mathbbm{I}_{v_{u,s_k}^{{\rm end}} \le 0}\ge\tau^{\rm{tr}}_{u,k} + i_{u,k}\frac{ d_{u, s_k} }{c}, \forall u \in \mathcal{U}, \forall k \in \mathcal{K},$}
\end{equation}
\begin{equation}\label{const:visibility_2}
\resizebox{.85\hsize}{!}{$v_{u,\hat{s}_k}^{{\rm end}} \,y_{u,k}\ge \sum_{\hat{k}\in\mathcal{K}} \left( \tau^{\rm{tr}}_{u,\hat{k}}
    + \tau^{\rm{prop,OP}}_{u,\hat{k}}
    + \tau^{\rm{comp}}_{u,\hat{k}} \right)
    + \tau^{\rm{prop,FP}}_{u,k} , \forall u \in \mathcal{U}, \forall k \in \mathcal{K},$}
\end{equation} 
\begin{equation}\label{const:visibility_3}
\resizebox{.85\hsize}{!}{$v_{u,\hat{s}_k}^{{\rm start}} \,y_{u,k}\le   \sum_{\hat{k}\in\mathcal{K}} \left( \tau^{\rm{tr}}_{u,\hat{k}}
    + \tau^{\rm{prop,OP}}_{u,\hat{k}}
    + \tau^{\rm{comp}}_{u,\hat{k}} \right),
\, \forall u \in \mathcal{U}, \forall k \in \mathcal{K},$}
\end{equation}
where $\mathbbm{I}_{v_{u,s_k}^{{\rm end}} \le 0} = 1$ if user $u$ is visible to satellite $s_k$ at transmission time and $0$ otherwise.
\section{Problem Formulation}
In this paper, our objective is to maximize the admitted tasks while promoting the effective use of computation and frequency resources. To achieve this, we consider the following objective function
\begin{equation}
\hspace{-.23 in}
\begin{split}
     \resizebox{.87\hsize}{!}{$\eta (\mathbf{I},\mathbf{Y},\mathbf{F},\mathbf{B}) =\sum_{u\in \mathcal{U}} \sum_{k\in \mathcal{K}} \ i_{u,k} - \alpha_1 \sum_{u\in \mathcal{U}} \frac{\sum_{k\in\mathcal{K}} \left(\rho_{u,k}^{tr} + \rho_{u,k}^{\rm comp} \right)}{\rho^{\rm max}_u} \vspace{1 in}$}\\ 
     \resizebox{.8\hsize}{!}{$-\alpha_2 \sum_{u\in \mathcal{U}} \frac{\sum_{k\in\mathcal{K}} \left( \tau^{\rm{tr}}_{u,k}
    + \tau^{\rm{prop,OP}}_{u,k} + \tau^{\rm{prop,FP}}_{u,k}
    + \tau^{\rm{comp}}_{u,k} \right)}{\tau_{u}^{\rm max}} $}, 
    \end{split}
\end{equation}
where the normalization factor $\rho^{\rm max}_u$ is the maximum power consumption for offloading task of user $u$. This function accounts for the number of admitted tasks while balancing between energy consumption and transmission delay by incorporating the weighting parameters \( \alpha_1, \alpha_2 \in [0,1]\). These parameters allow for adjustable emphasis on energy efficiency versus delay minimization, depending on system requirements. The final problem is formulated as follows:
\allowdisplaybreaks
\begin{subequations}
\label{eq:P1}
\begin{flalign}
\centering
\mathcal{P}_1: \quad& \max_{\mathbf{I},\mathbf{Y},\mathbf{F},\mathbf{B}} \eta(\mathbf{I},\mathbf{Y},\mathbf{F},\mathbf{B}) \label{P1:Objective}\\
\text{s.t.}  \quad&\eqref{const:offloading_i}- \eqref{const:offloading_closed_path},\eqref{const:Frequency_f_sum},\eqref{const:i_b_relation},\eqref{const:Computing_sum}, \eqref{const:computing_relation},
\eqref{const:Delay}- \eqref{const:visibility_3}, \label{P1:const_1}\\
&0\leq b_{u,k,s_k}\leq 1 ,\forall u \in {\mathcal{U}}\, , \,\forall k \in \mathcal{K},  \,\forall s_k \in \mathcal{S}, \label{P1:const_2}\\
&0\leq f_{u,k,\hat{s}_k}\leq 1  ,\forall u \in {\mathcal{U}}\, , \,\forall k \in \mathcal{K},  \,\forall s_k \in \mathcal{S}, \label{P1:const_3}\\
& i_{u,k},y_{u,k} \in\{0,1\},\,\forall u \in {\mathcal{U}}\,  \forall k \in \mathcal{K}. \label{P1:const_4}
\end{flalign}
\end{subequations}
 Problem \(\mathcal{P}_1\) is a non-convex mixed-integer non-linear programming (MINLP) problem, making it challenging to solve directly. Particularly, objective \eqref{P1:Objective} and constraints \eqref{const:Delay}$-$\eqref{const:visibility_3} are not convex due to the term $\frac{1}{x}$ which is not convex for $x\ge 0$, where $x$ represents $b_{i,k,s_k}$ and $f_{i,k,\hat{s}_k}$.
\section{Solution approach}
In this section, we present our methodology for (mixed-integer) convexifying the non-convex objective and constraints of  $\mathcal{P}_1$. We refer to a problem as mixed-integer convex when relaxing its integer variables to continuous ones yields a convex optimization problem. We begin by introducing auxiliary variables $\vartheta_{u,k,s_k}$ and $\varphi_{u,k,\hat{s}_k}$  representing the fraction of bandwidth and frequency resources allocated to the $u$-th user in the $k$-th path at both access and destination nodes respectively, which  can be written as:  
 \begin{equation} \label{const: vartheta_1}
 \vartheta_{u,k,s_k}\ge\frac{1}{b_{u,k,s_k}} , \forall u \in \mathcal{U},\forall k \in \mathcal{K},\forall s_k \in \mathcal{S},
 \vspace{-.05 in}
\end{equation}
 \begin{equation} \label{const: varphi_1}
\varphi_{u,k,\hat{s}_k}\ge\frac{1}{f_{u,k,\hat{s}_k}} , \forall u \in \mathcal{U},\forall k \in \mathcal{K},\forall \hat{s}_k \in \mathcal{S}. 
\vspace{-.07 in}
\end{equation}
 \begin{equation} \label{const: varphi_vartheta}
\hspace{0.35 in}\varphi_{u,k,\hat{s}_k},  \vartheta_{u,k,s_k}\ge 0 , \forall u \in \mathcal{U},\forall k \in \mathcal{K},\forall \hat{s}_k \in \mathcal{S}. 
\end{equation}
Then, transmission and computing delays are expressed as 
\begin{equation}\label{eq:transmission_time_updated} 
\hspace{-0.0 in}\tau^{\rm{tr}}_{u,k}= \frac{\zeta_u \, \vartheta_{u,k,s_k}}{\,{\rm{B}}_{s_k}\,{\rm r}_{u,s_k}}, \forall u\in \mathcal{U}, \forall k\in \mathcal{K},
\end{equation}
\begin{equation}  \label{eq:computation_time_1}    \tau^{\rm{comp}}_{u,k}=  \frac{X_u\, \varphi_{u,k,\hat{s}_k}}{ \rm{F}_s}, \forall u\in \mathcal{U}, \forall k\in \mathcal{K}. 
\end{equation}

By introducing these auxiliary variables, we reformulate the original non-convex constraints, namely constraints \eqref{const:Delay}$-$\eqref{const:visibility_3}, into upper bound convex (affine) forms. Consequently, the objective function becomes concave, as confirmed by the Hessian matrix, which is negative semidefinite. Specifically, the second derivative is \(-2 \kappa\, X_u \rm{F}_x^2\) with respect to \(\mathbf{F}\) and zero otherwise. The definiteness of the Hessian matrix ensures that the reformulated problem has a concave objective function, making its maximization tractable. However, constraints \eqref{const: vartheta_1} and \eqref{const: varphi_1} remain non-convex due to the potential division by zero for frequency and computing resources allocation, which occurs if a task is not offloaded. To address this issue, we adopt disjunctive (Big-M) constraints as follows 
\begin{equation}\label{const:vartheta_2}
\vartheta_{u,k,s_k}\le {\rm M}_1  i_{u,s}, \forall u \in \mathcal{U}, \forall k \in \mathcal{K},  \forall s_k \in \mathcal{S},
\end{equation}
\begin{equation}
 \resizebox{.85\hsize}{!}{$\vartheta_{u,k,s_k}\ge \frac{1}{b_{u,k,s_k}+\epsilon_1}  - ( 1 -i_{u,k})  {\rm M}_1, \forall u \in \mathcal{U}, \forall k \in \mathcal{K},  \forall s_k \in \mathcal{S}, $}
\end{equation}
\begin{equation}
\varphi_{u,k,\hat{s}_k}\le {\rm M}_2  i_{u,k},  \forall u \in \mathcal{U}, \forall k \in \mathcal{K},  \forall \hat{s}_k \in \mathcal{S}, 
\end{equation}
\begin{equation}\label{const:varphi_3}
 \resizebox{.85\hsize}{!}{$\varphi_{u,k,\hat{s}_k}\ge \frac{1}{f_{u,k,\hat{s}_k}+\epsilon_2}  - ( 1 -i_{u,k}){\rm M}_2,\,   \forall u \in \mathcal{U}, \forall k \in \mathcal{K},  \forall \hat{s}_k \in \mathcal{S}.$} 
\end{equation}
Here, ${\rm M}_1$ and ${\rm M}_2$ denote a sufficiently large constants, while $\epsilon_1$  and $\epsilon_2$ are a very small positive number. These parameters are chosen to ensure that   $\vartheta_{u,k,s_k} = 0$ and $\varphi_{u,k,\hat{s}_k} = 0$ when the task is not offloaded and greater than zero if the task is offloaded. Finally, problem $\mathcal{P}_1$ can be written as, 
\begin{subequations}
\label{eq:P2}
\begin{flalign}
\centering
\mathcal{P}_2: & \max_{\mathbf{I}, \mathbf{Y},\mathbf{F},\mathbf{B}, \boldsymbol{\vartheta},\boldsymbol{\varphi}} \eta (\mathbf{I},\mathbf{Y},\mathbf{F},\mathbf{B}, \boldsymbol{\vartheta},\boldsymbol{\varphi})  \label{P2:Objective}\\
\text{s.t.} \quad&\eqref{P1:const_1}- \eqref{P1:const_4},\eqref{const: varphi_vartheta},
\eqref{const:vartheta_2}- \eqref{const:varphi_3}.
\end{flalign}
\end{subequations}
Problem $\mathcal{P}_2$ is convex 
mixed-integer non-linear program (MINLP), allowing us to use various off-the-shelf tools, such as \textit{cvxpy} \cite{diamond2016cvxpy} and \textit{Gurobi} \cite{gurobi}. The overall complexity depends on the chosen solver and algorithm \cite{8716508}.
\section{Evaluation of the Proposed Architecture}
In this section, we present comprehensive numerical results to assess the performance of the proposed offloading scheme using a Python-based simulator. Specifically, the \textit{Skyfield} package \cite{rhodes2019skyfield} is employed to model the deployment of LEO satellites, simulating the Starlink constellation through the adoption of the latest two-line element (TLE) set \cite{celestrak}. This data enables accurate estimation of the position and velocity of Starlink satellites. Additionally, the simulation includes $30$ MEO satellites positioned at an altitude of $2100$~km, distributed across three orbital planes with inclination angles of $86^{\circ}, 88^{\circ},$ and $90^{\circ}$, forming near-polar orbits. The MEO satellites are deployed to enhance coverage in the northern region of Québec, Canada, particularly around Ivujivik (latitude $62^\circ$, longitude $-77.5^\circ$). The LEO and MEO satellites are also assumed to have $25$ and $50$ frequency resources, respectively, with each resource block of \SI{180}{\kilo\hertz}. The computational capabilities of the LEO and MEO satellites are set to $5$~GHz and $10$~GHz, respectively \cite{9515574}. The effective switch capacitance is set at $\kappa=10^{-28}$~ \({\text{Watt}\cdot \text{sec}^3}/{\text{cycle}^3}\). User transmission power is assumed to be $17$~dBm with an antenna gain of $5$~dBi, while the antenna gain for both LEO and MEO satellites is set at $20$~dBi \cite{9515574}.

The simulation also considers a set of $100$ users, randomly distributed within the area surrounding the target zone. The user-to-satellite channels are modelled by large-scale fading based on free-space path loss and small-scale fading, which follows a Rician distribution with a Rician factor of $10$ \cite{9515574}. Task sizes are randomly select a  $1$ or $2$~Mbits and $\tau^{\rm th }_u =3$~sec. The required processing cycles is $1000$~cycles$/$bit The background noise density is set to $-174 \, \text{dBm/Hz}$ \cite{9515574}. The optimization parameters $\alpha_1$ and $\alpha_2$ are both set to $0.1$ to penalize packet blocking rates while balancing task delay and energy consumption. The simulation results are based on $500$ independent Monte Carlo realizations where a subset of users is active following Poisson distribution with rate $\lambda=5$. Because we focus on remote areas with limited terrestrial network availability, we consider two baseline scenarios in addition to the proposed {hybrid MEO-LEO network}: (1) a {LEO-only} network for communication and task offloading and (2) a standalone MEO ({MEO-only}) network.
\begin{figure}[t!]
\pdfimageresolution=300
\centering
\begin{subfigure}[b]{1\columnwidth}
\centering
\includegraphics[width=1\columnwidth,draft=false]{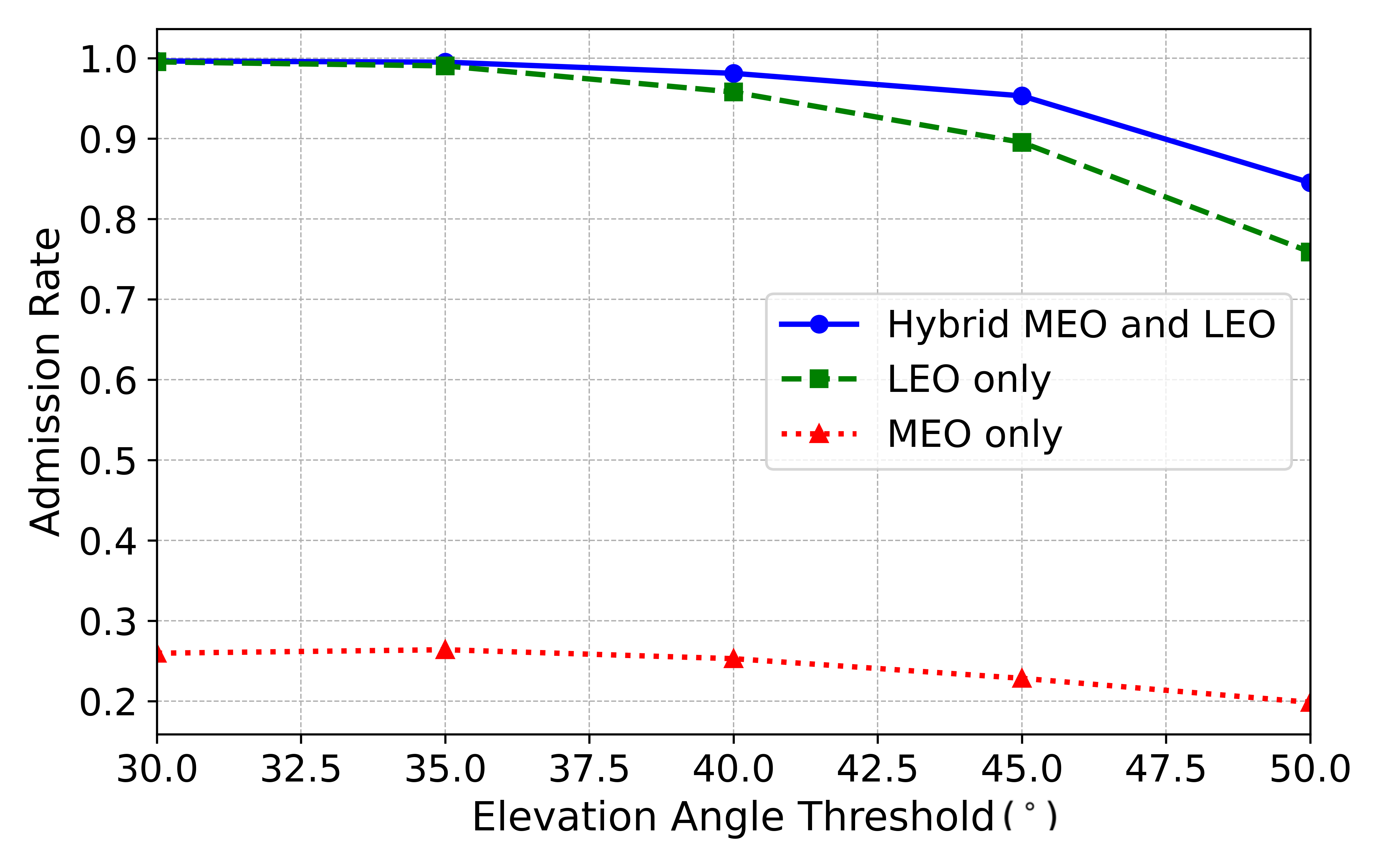}
\caption{Admission rate vs elevation angle.}
\label{Fig:Admission rate}
\end{subfigure}\\~\vspace{-.0 in}
\begin{subfigure}[b]{1\columnwidth}
\centering
\includegraphics[width=1\columnwidth,draft=false]{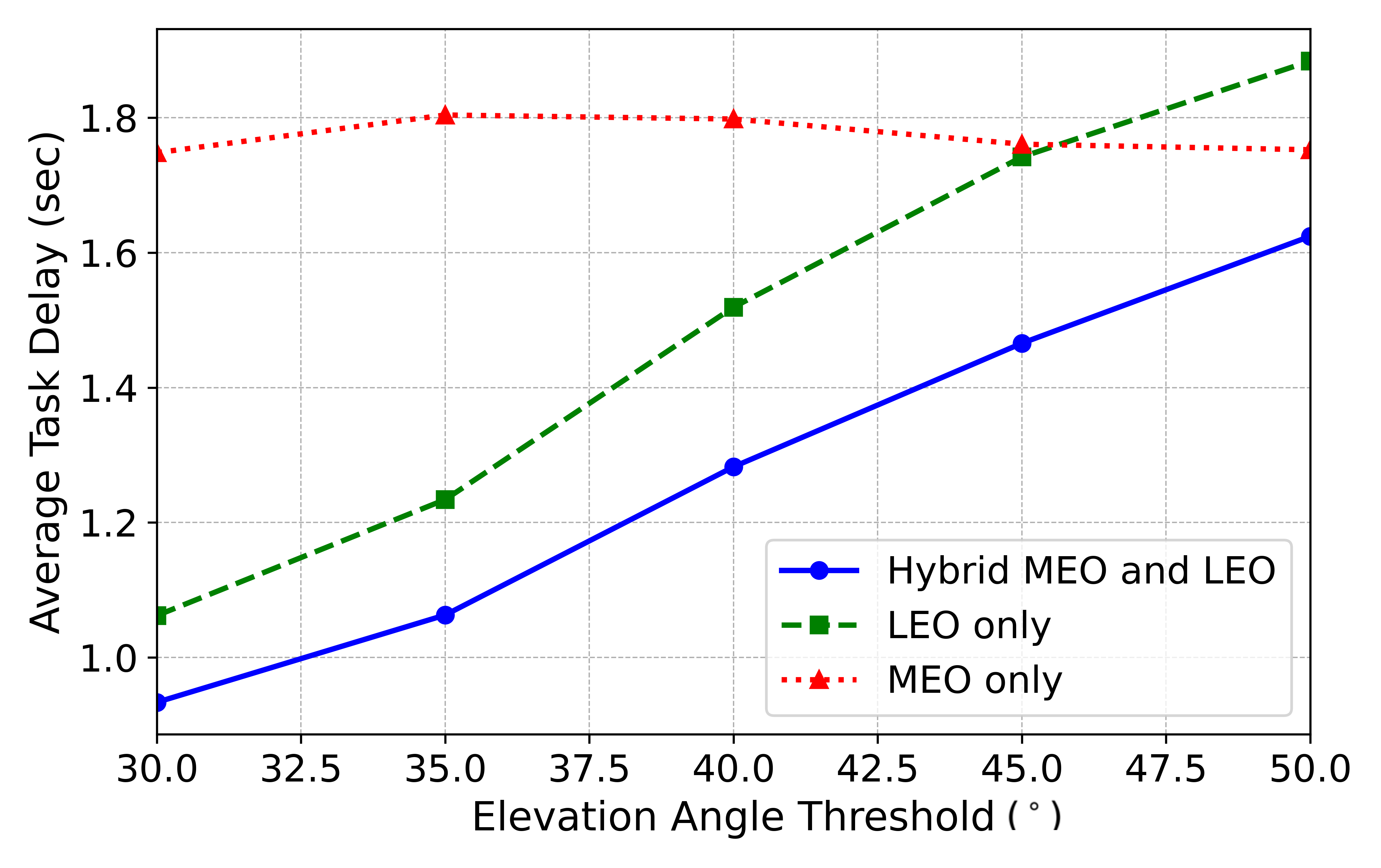}
\caption{Average delay vs elevation angle.  }
\label{Fig:Delay}
\end{subfigure}\\~\vspace{-.0 in}
\begin{subfigure}[b]{1\columnwidth}
\centering
\includegraphics[width=1\columnwidth,draft=false]{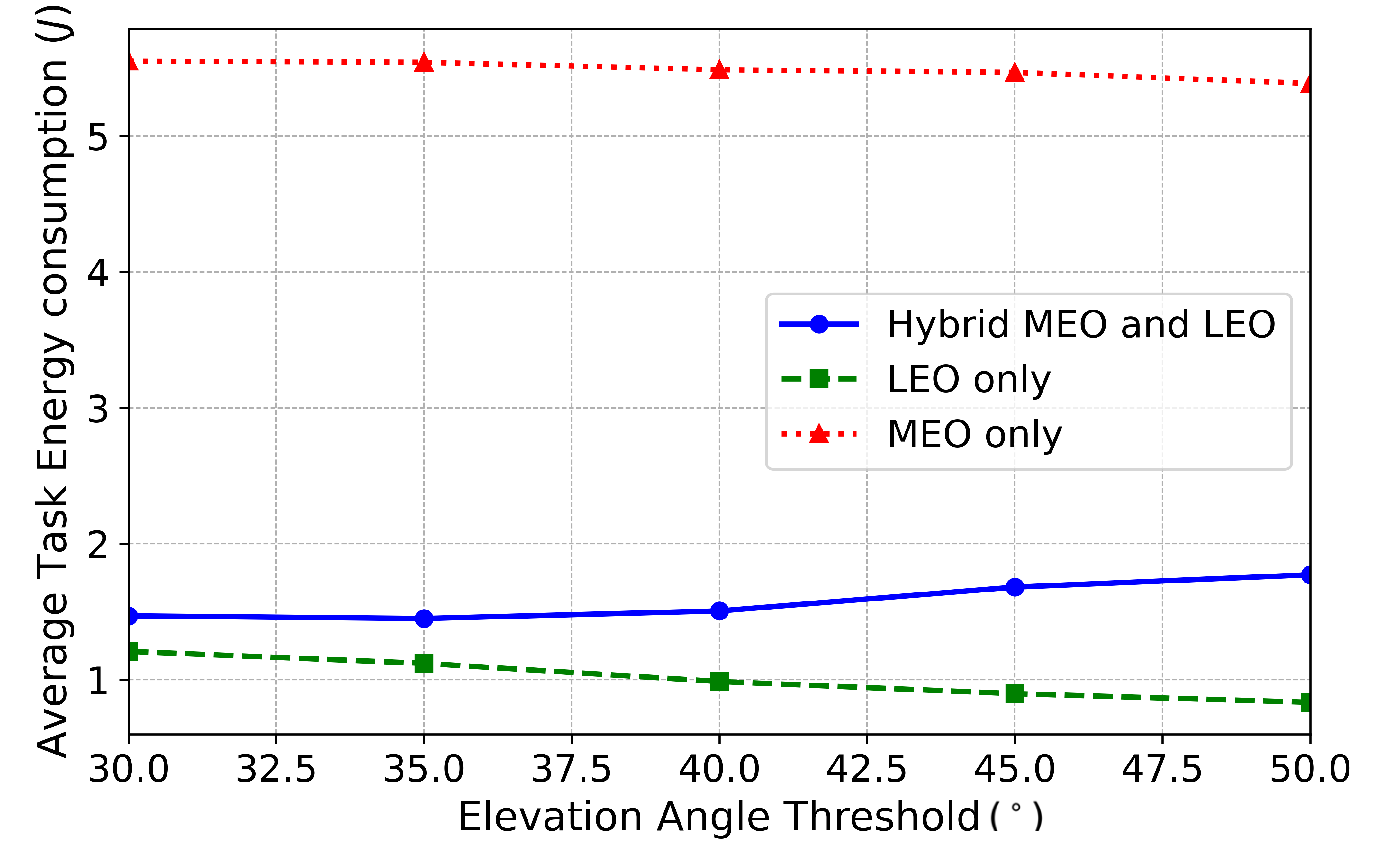}
\caption{Average energy consumption vs elevation angle.  }
\label{Fig:Energy}
\end{subfigure}
\caption{ Numerical performance analysis. }
\label{fig:Performance evaluation}
\vspace{-.35in}
\end{figure}

Fig. \ref{fig:Performance evaluation} presents the results of the proposed hybrid LEO-MEO architecture compared to using either MEO or LEO satellite networks independently. The task admission rate against the elevation angle threshold is shown in Fig. \ref{Fig:Admission rate}. It is evident that the proposed hybrid LEO-MEO offloading scheme outperforms the baseline approaches, with the performance gain increasing as the elevation angle threshold rises, reaching up to $15\%$ at an elevation angle threshold of $50^\circ$ degrees compared to the LEO network. This can be justified by the decrease in the number of visible LEO satellites as the elevation angle threshold increases. However, the number of visible MEO satellites decreases slightly due to their higher altitudes, which maintain higher elevation angles, which is also shown in the performance of the MEO network. Despite this, the hybrid architecture provides additional communication and computational paths, allowing more tasks to be admitted. It is also clear that using only LEO satellites results in a higher admission rate compared to using only MEO satellites. This is due to the higher path loss in MEO satellites compared to LEO networks, which limits the number of offloaded tasks given MEOs' limited frequency and computational resources.

Figs. \ref{Fig:Delay} and \ref{Fig:Energy} illustrate the average task delay and energy consumption against the elevation angle threshold. For the proposed 
scheme, the average delay and the energy consumption increase as the elevation angle threshold increases. This is because of the increased number of tasks offloaded to MEO satellites with higher transmission energy consumption. This is further supported by the behaviour of the MEO network, indicating that energy consumption is dominated by MEO transmission energy. Additionally, the LEO network experiences higher delay compared to the proposed hybrid scheme, up to $12\%$ at $50^\circ$, as more tasks are allocated to fewer LEO satellites, resulting in increased transmission, propagation, and computational delays. This effect becomes clearer as the elevation angle increases, reducing the number of visible LEO satellites. It is worth noting that the MEO network exhibits a decrease in average delay as fewer tasks are admitted due to limited visible MEO satellites, leading to better utilization of computational resources.
\section{Conclusion}
This paper addresses the task offloading problem in a hybrid LEO-MEO architecture to maximize the task admission rate while balancing energy consumption and delay, given specific delay requirements. As the formulated problem is a mixed-integer nonlinear program (MINLP), we apply disjunctive constraints and convex relaxation techniques to reformulate the problem into a convex form, which is readily solved using standard optimization tools. Through extensive simulations, we demonstrate that the proposed hybrid LEO-MEO architecture achieves an increase in admission rate by up to $15\%$ and a reduction in average delay by up to $12\%$ at an elevation angle threshold of $50^\circ$, compared to a LEO-only satellite network. These results underscore the potential of the proposed architecture to enhance service reliability in Arctic regions.

\bibliographystyle{IEEEtran}
\bibliography{IEEEabrv,Bibliography}

\end{document}